\documentclass[10pt]{article}
\usepackage{epsfig}
\usepackage{epstopdf}
\usepackage{lineno}

\newcommand{\vl}{V_{\rm L}}
\newcommand{\vr}{V_{\rm R}}
\newcommand{\gl}{g_{\rm L}}
\newcommand{\gr}{g_{\rm R}}

\newcommand{\fp}{F_{\rm R}}
\newcommand{\fm}{F_{\rm L}}
\newcommand{\fz}{F_{\rm 0}}

\def\beq{\begin{equation}}
\def\eeq#1{\label{#1}\end{equation}}
\def\eeqn{\end{equation}}

\def\beqa{\begin{eqnarray}}
\def\eeqa#1{\label{#1}\end{eqnarray}}
\def\eeqan{\end{eqnarray}}

\let\bar=\overbar

\def\Dslash{\not{\hbox{\kern-4pt $D$}}}
\def\dslash{\not{\hbox{\kern-2pt $\del$}}}

\def\msb{{\bar{\ssstyle M \kern -1pt S}}}

\def\Title#1{\begin{center} {\Large {\bf #1} } \end{center}}

\begin{document}
%\linenumbers

\Title{Searches for new physics in top decays at the LHC}

\bigskip\bigskip

%+\addtocontents{toc}{{\it D. Reggiano}}
%+\label{ReggianoStart}

\begin{raggedright}  

{\it Antonio Onofre \index{Onofre, A.}, On behalf of the ATLAS and CMS Collaborations \\[2mm]
Departamento de F\'{\i}sica, \\
Universidade do Minho \\
Campus de Gualtar, \\
4710 - 057 Braga, Portugal. \\[2mm]
Proceedings of CKM 2012, the 7$^{th}$ International Workshop on the CKM Unitarity Triangle, 
University of Cincinnati, USA, 28 September - 2 October 2012}
\bigskip\bigskip
\end{raggedright}

\section{Abstract}

The search for new physics in top quark decays at the LHC is reviewed in this paper. Results from ATLAS~\cite{Aad:2008zzm} and CMS~\cite{Chatrchyan:2008aa} 
experiments on top quark decays within the Standard Model are presented together with the measurements of the W boson 
polarizations and the study of the structure of the $Wtb$ vertex. As a natural step forward, the experimental status on
measurements sensitive to top quark couplings to gauge bosons ($\gamma$, Z, W and H) is reviewed as well as possible
top quark decays Beyond the Standard Model (MSSM and FCNC).

\section{Introduction}

The top quark is the heaviest known quark ($m_t=173.18 \pm 0.94$~GeV~\cite{Aaltonen:2012ra}) ever discovered. With a charge of $+2/3|e|$ ($e$ is the 
electron charge) and spin $1/2$, the top quark completes the 3 family structure of the Standard Model (SM). One of the stricking features of this 
quark is precisely the value of its mass, so distinctive from the other fermions, which may suggest it plays a more fundamental role in the 
eletroweak symmetry breaking mechanism. In the SM, top quarks are expected to decay mostly through $t\to bW$, being the branching ratios 
$Br(t\to sW)\le$~0.18\% and $Br(t\to dW)\le$~0.02\% diminut when compared with the dominant decay channel~\cite{PDG}. The lifetime, at least one order of 
magnitude smaller than the typical hadronization time scale $O(10^{-23})$~s, was measured to be $\tau_t=(3.29^{+0.90}_{-0.67})\times 10^{-25}$~s 
\cite{Abazov:2012vd}. This implies top quarks decay before hadronization may take place and spoil the spin information propagated to the final state 
decay products. Once again, top quarks show remarkable distintive features when compared to the other quarks. At the LHC, top quarks are mainly 
produced in pairs through gluon fusion or quark anti-quark annihilation with a predicted cross section of $164.6^{+11.4}_{-15.7}$~pb (approx. NNLO) 
at 7~TeV \cite{Aliev:2010zk}. There is however a significant number of top quarks which are produced singly through the $t$-channel ($gq'\to q t\bar 
b$), associated production ($bg\to Wt$) and $s$-channel ($q\bar q'\to t\bar b$) single top production. The production cross sections are, 
respectively, $64.6^{+2.7}_{-2.0}$~pb~\cite{Kidonakis:2011wy} (NNLO+soft gluon corrections), $15.7\pm 1.1$~pb~\cite{Kidonakis:2010ux} (approx. NNLO) 
and $4.6 \pm 0.2$~pb~\cite{Kidonakis:2010tc} (NNLL resummation). In this paper the experimental status of top quark measurements sensitive to gauge 
bosons couplings (W, $\gamma$, Z and H) and decays beyond the SM are reviewed. As most analysis described in the paper rely on the lepton+jets and 
dilepton final states, or variations from these, it is appropriate to shortly mention them.

In the lepton+jets channel, one isolated lepton ($e$ or $\mu$) is required in the event selection as well as at least 4 jets for both 
ATLAS~\cite{Aad:2012qf,Aad:2012hg} and CMS~\cite{Chatrchyan:2011yy,:2012qka}. While for muons, the transverse momentum ($p_{\rm T}$) selection is similar for both 
experiments, selected electrons at ATLAS are harder (with transverse energy, $E_{\rm T}$, above 25~GeV) than in CMS ($E_{\rm T}>$~20~GeV). Softer jets, with 
$p_{\rm T}>$~25~GeV, are also selected by ATLAS compared to CMS ($p_{\rm T}>$~30~GeV). The use of a transverse missing energy ($E_{\rm T}^{\rm miss}$) cut 
together with the requirement that at least one jet had to be tagged as a $b$-jet was performed by ATLAS while CMS required at least two $b$-tagged jets with no 
$E_{\rm T}^{\rm miss}$ condition.

In the dilepton channel, events with two isolated leptons ($e$ or $\mu$) and two jets, from which at least one is $b$-tagged, are selected by both 
CMS~\cite{:2012bta,:2012qka} and ATLAS~\cite{ATLAS:2012aa}. While at least 2 opposite charge leptons are requested by CMS, exactly 1 pair is imposed by 
ATLAS. In order to reduce potential contributions from the $Z$+jets background, events are required to have the invariant mass of the two opposite charge 
leptons ($m_{\ell^+ \ell^-}$) outside a mass window of at least 10~GeV around the central value of the $Z$ mass. Both CMS and ATLAS impose $E_{\rm 
T}^{\rm miss}$ cuts to events and ATLAS, for the $e^{\pm}\mu^{\mp}$, requires the scalar sum of jets and leptons transverse energies ($H_{\rm T}$) to be 
above 130~GeV.

\section{The $Wtb$ vertex structure}

\begin{figure}[htb]
\begin{center}
\epsfig{file=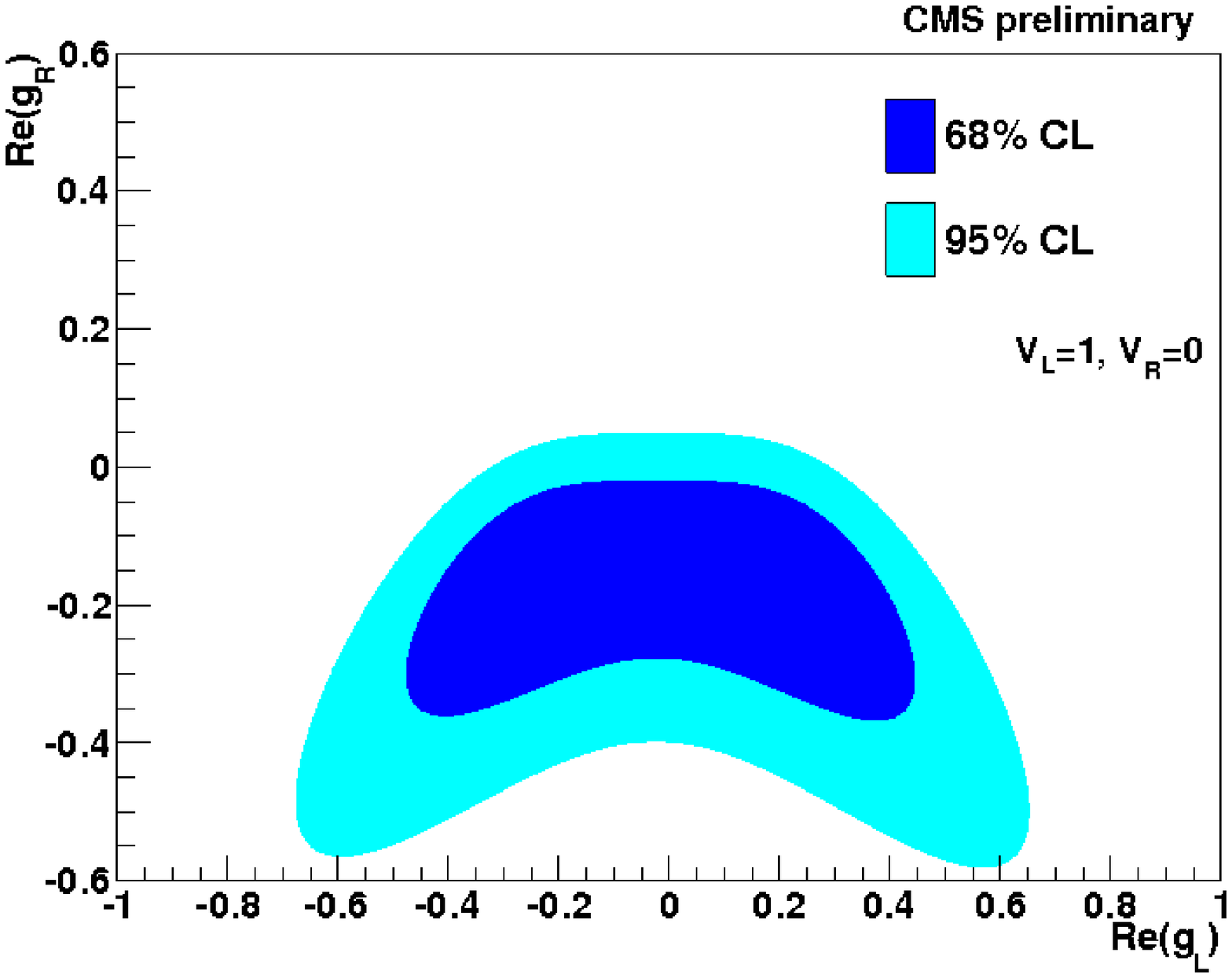,height=2.in} \epsfig{file=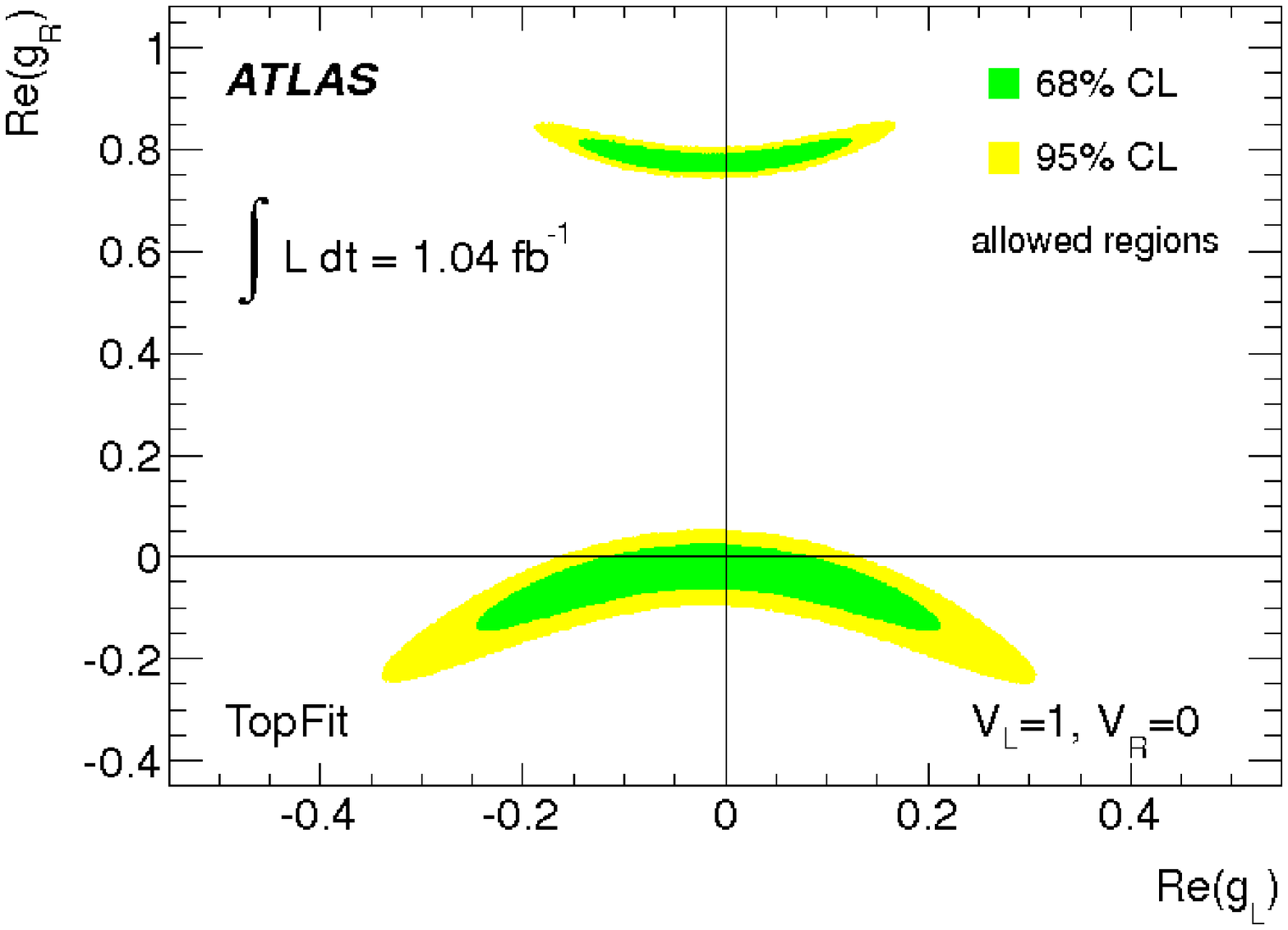,height=2.in}
\caption{CMS(left) and ATLAS(right) allowed regions, at 68\% and 95\% CL, on the anomalous coupling plane ($\gl$,$\gr$), for $\vr$=0 and $\vl$=1.}
\label{fig:gLgR}
\end{center}
\end{figure}

In the SM, the Wtb vertex has a $(V-A)$ structure where, V and A, are the vector and axial-vector contributions to the vertex. One way of probing the structure 
of the vertex, is to study top quark decays to $W$-bosons and $b$-quarks ($t\to bW$). The W bosons produced in these decays can have longitudinal, left-handed or 
right-handed polarizations with fractions $F_0$, $F_{\rm L}$ and $F_{\rm R}$, respectively equal to $0.687\pm 0.005$, $0.311\pm0.005$ and $0.0017\pm0.0001$ at 
NNLO~\cite{Wtb:NNLO}, in the SM. The helicity fractions were extracted from a direct fit to the top quark decay products angular distribution, $\cos \theta$, using 
templates from simulation. Alternatively, angular asymmetries ($A_{+}$, $A_{-}$ and $A_{FB}$), built from $\cos \theta$, were also used by ATLAS~\cite{Wtb:Asym}. 
The angle $\theta$¸ is measured between the momentum direction of the charged lepton from the decay of the $W$ boson, and the reversed momentum direction of the 
$b$-quark from the decay of the top quark, both boosted into the $W$ boson rest frame. CMS~\cite{CMS-PAS-TOP-11-020} measured the helicity fractions to be $\fz$ = 
0.567$\pm$0.074(stat)$\pm$0.047(syst), $\fm$ = 0.393$\pm$0.045(stat)$\pm$0.029(syst) and $\fp$ = 0.040$\pm$0.035(stat)$\pm$0.044(syst). The correspondent 
measurements from ATLAS~\cite{Aad:2012ky} are, $\fz$ = 0.67$\pm$0.03(stat)$\pm$0.06(syst), $\fm$ = 0.32$\pm$0.02(stat)$\pm$0.03(syst) and $\fp$ = 
0.01$\pm$0.01(stat)$\pm$0.04(syst). All these measurements are consistent with SM expectations. The measured fractions were used to probe the existence of anomalous 
$Wtb$ couplings. Exclusion limits on the real components of the anomalous couplings $\gl$ and $\gr$ were set at 68\% and 95\% CL (Figure~\ref{fig:gLgR}). The region 
of $\gr$ around 0.8 in the ATLAS plot is disfavored by the current experimental measurements on the single top quark production cross sections at Tevatron and LHC. 
Once the limits depend on $V_{\rm tb}$, it is worth mentionning the current experimental status of these measurements (Table~\ref{tab:Vtb}) which assume the SM and 
negligeable decays to $t\to sW$ and $t\to dW$, when compared to the dominant channel. The current $V_{\rm tb}$ measurements precision is at the level of 5\%-10\%.

%%%%%%%%%%%%%%%%%%%%%%%%%%%%%%%%%%%%%%%%%%%%%%%%%%%%%%%%%%%%%%%%%%%%%%%%%%%
%%%%%%%%%%%%%%%%%%%%%%%%%%%%%%%%%%%%%%%%%%%%%%%%%%%%%%%%%%%%%%%%%%%%%%%%%%%

%%%%%%%%%%%%%%%%%%%%%%%%%%%%%%%%%%%%%%%%%%%%%%%%%%%%%%%%%%%%%%%%%%%%%%%%%%%
\begin{table}[b]
\begin{center}
{\scriptsize
\begin{tabular}{|c|c|}
 \hline
 {\bf ATLAS}                                                    &               {\bf Measurement (SM)}                  \\
    { $t$-channel (@~7TeV)}                    & {  $|V_{tb}|=1.13^{+0.14}_{-0.13}$ }                 \\
    { (arXiv:1205.3130)   }                    & {  $|V_{tb}|\ge$0.75 @ 95\% CL ($|V_{tb}$ in [0,1])]}          \\[0.5mm]
    { $t$-channel (@~8TeV,5.8~fb$^{-1}$)}     & { $|V_{tb}|=1.04^{+0.10}_{-0.11}$ }                         \\
    { (ATLAS-CONF-2012-132)}                  & { $|V_{tb}|\ge$0.80 @ 95\% CL ($|V_{tb}$ in [0,1])]}          \\[0.5mm]
    { $Wt$-prod.  (PLB 716~(2012)~142)}        & {  $|V_{tb}|=1.03^{+0.16}_{-0.19}$}                          \\
 \hline
 {\bf CMS}                                                      &        {\bf Measurement (SM)}         \\
   {  $t$-channel (@~7TeV)}                    &  {  $|V_{tb}|$=1.02$\pm$0.046(exp)$\pm$0.017(th.)}   \\[0.5mm]
   { (arXiv:1209.4533)}                        &  {  $|V_{tb}|\ge$0.92 @ 95\% CL ($|V_{tb}$ in [0,1])]}         \\[0.5mm]
   { $t$-channel (@~8TeV,5.~fb$^{-1}$)}       &  { $|V_{tb}|$=0.96$\pm$0.08(exp)$\pm$0.02(th.)}     \\
   { (CMS PAS TOP-12-011)}                    &  { $|V_{tb}|\ge$0.81 @ 95\% CL ($|V_{tb}$ in [0,1])]}         \\[0.5mm]
   {    $Wt$-prod.  (@~7TeV)}                  &  {  $|V_{tb}|=1.01^{+0.16}_{-0.13}$(exp.)$^{+0.03}_{-0.04}$(th.)}  \\
   { (arXiv:1209.3489)}                        &  {  $|V_{tb}|\ge$0.79 @ 95\% CL ($|V_{tb}$ in [0,1])]}         \\[0.5mm]
 \hline
\end{tabular}
}
\caption{Current status of $V_{\rm tb}$ measurements from both ATLAS and CMS.}
\label{tab:Vtb}
\end{center}
\end{table}
%%%%%%%%%%%%%%%%%%%%%%%%%%%%%%%%%%%%%%%%%%%%%%%%%%%%%%%%%%%%%%%%%%%%%%%%%%%

\section{Top quark couplings to gauge bosons}

\begin{figure}[htb]
\begin{center}
  \begin{tabular}{ccc}
    \hspace*{-1cm}\epsfig{file=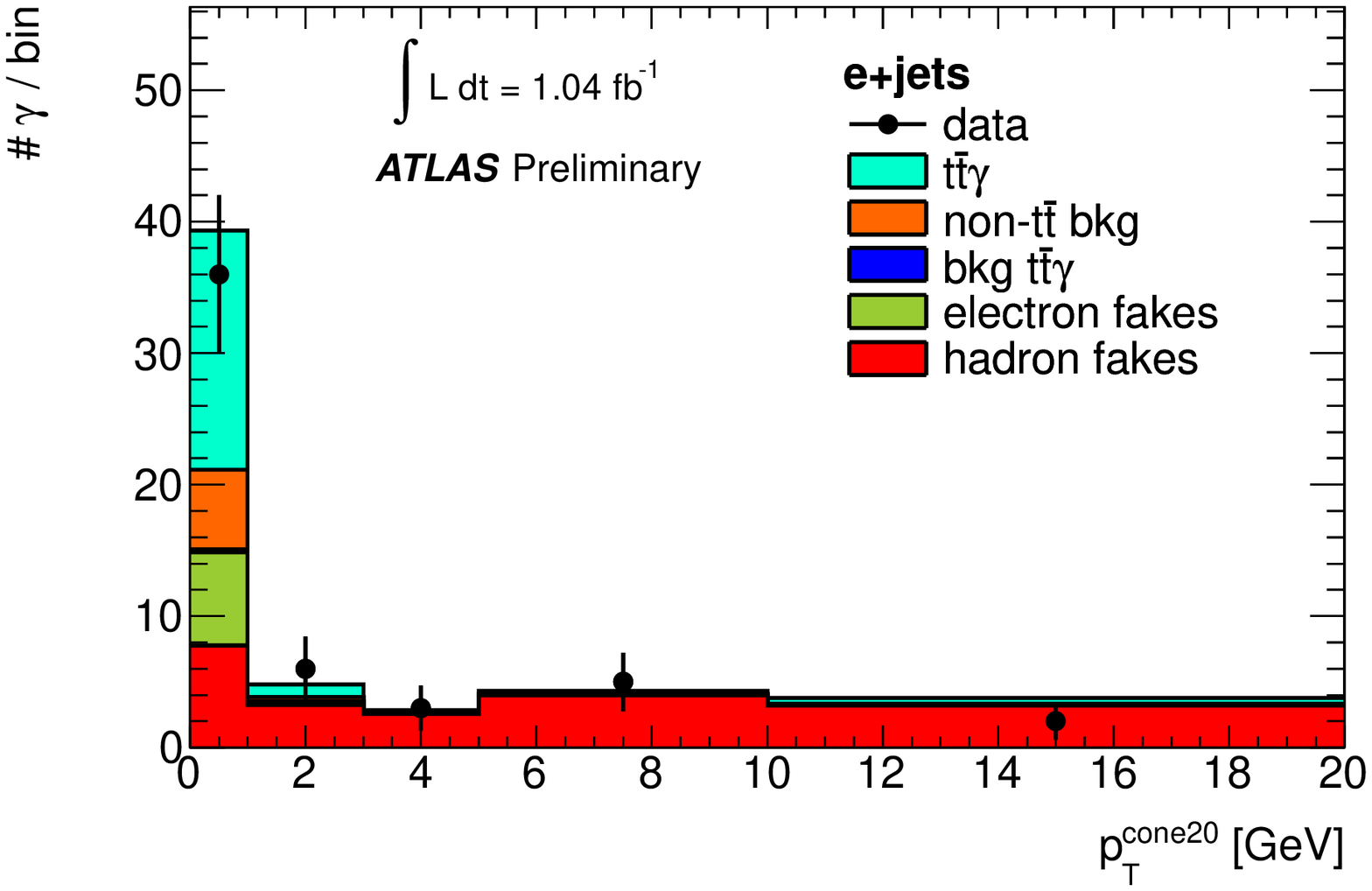,height=1.65in} & \epsfig{file=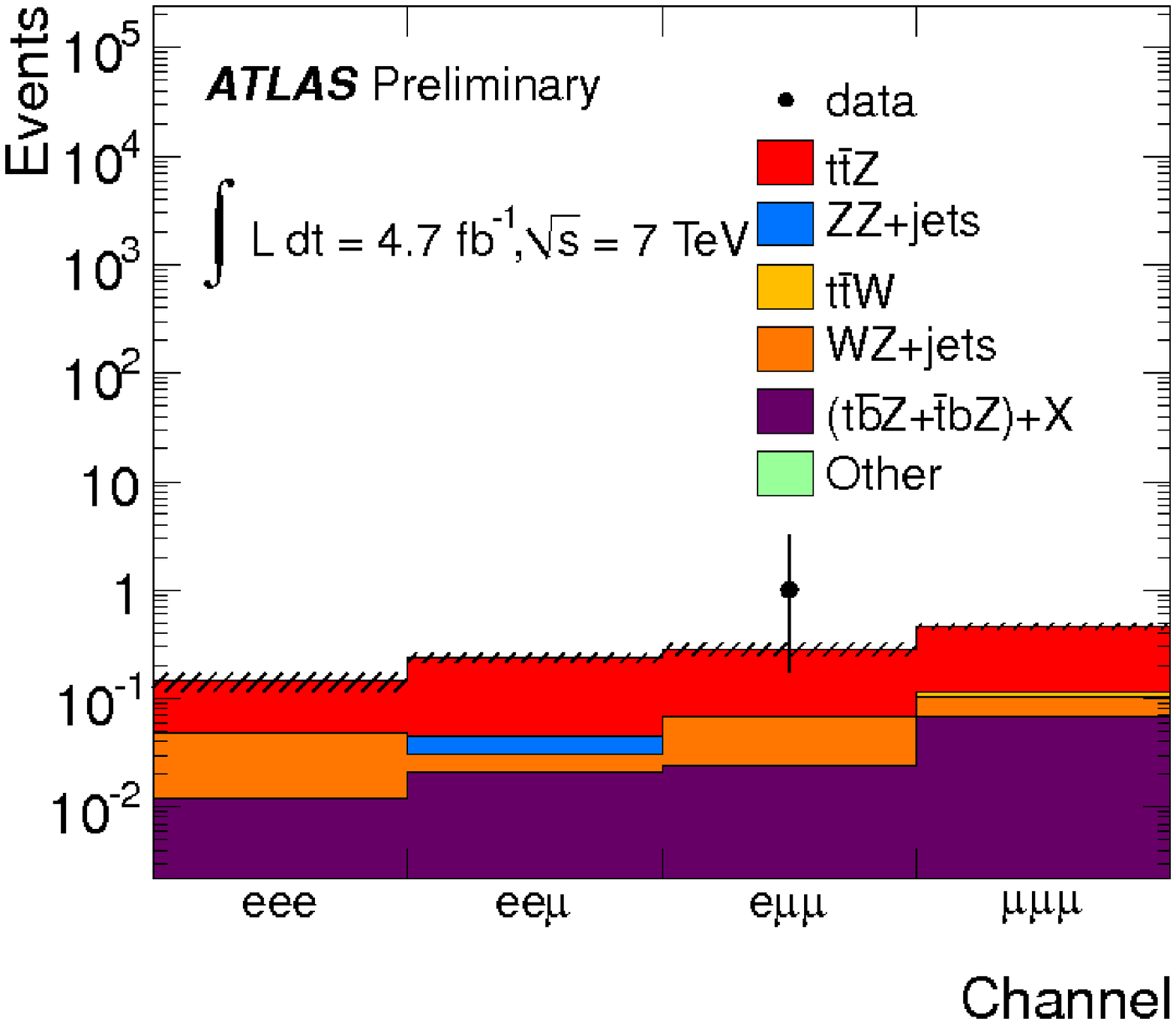,height=1.65in} & \epsfig{file=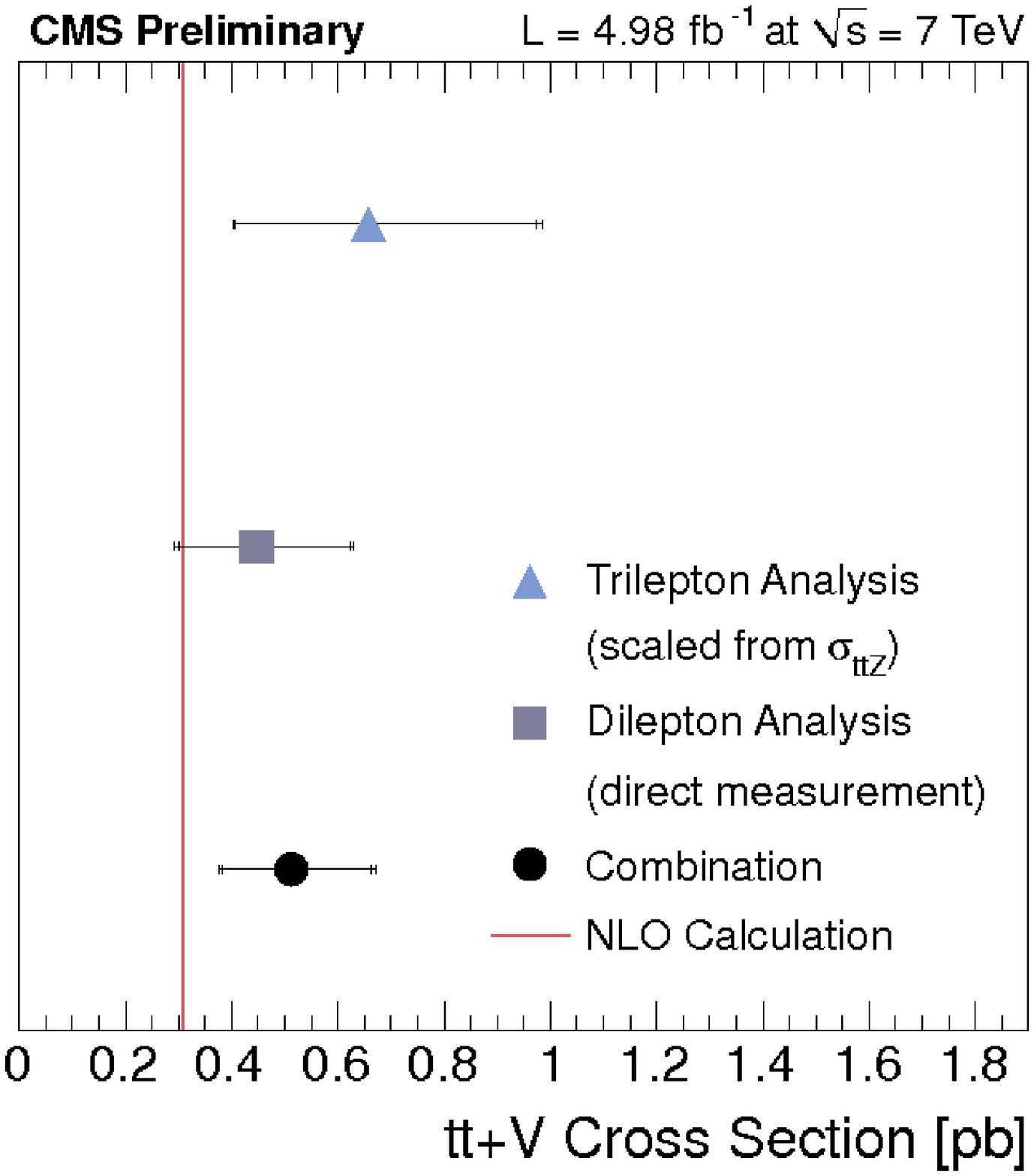,height=1.65in} \\
  \end{tabular}
\caption{(left) ATLAS template fit to $t\bar t\gamma$ events in the electron channel; (middle) ATLAS expected and observed numbers of events for the $t\bar t Z$ signal 
region; (right) CMS measurements of the $t\bar t V$ production cross section.}
\label{fig:ttGaugeBosons}
\end{center}
\end{figure}

Following the discussion on the $Wtb$ vertex and $V_{\rm tb}$, it is natural to ask what is the current LHC status on the measurements of the 
couplings of the top quarks to other bosons ($\gamma$,Z,W and H). Several dedicated analyses were developed by both ATLAS and CMS, relying to a great 
extent on the lepton+jets and dilepton channels discussed previously, with the exception that activity from the gauge bosons is expected to be 
present as well in the events. These are particularly challenging channels which should be improved as both CMS and ATLAS collect more data over time. 
ATLAS~\cite{ATLAS-CONF-2011-153} looked into $t\bar t \gamma$ production and, by fitting the data with templates for prompt photons (well 
isolated) and hadron fakes, measured the cross section to $\sigma(t\bar t \gamma) = 2.0 \pm 0.5 ~{\rm(stat)} \pm 0.7 ~{\rm(syst)} \pm 0.08 
~{\rm(lumi.)}$~pb, for events with photons with $p_{\rm T}>$8~GeV for which the expected SM cross section is 2.1$\pm$0.4~pb (at 
7~TeV). For the $t\bar t Z$ channel ATLAS~\cite{ATLAS-CONF-2012-126} observes 1 candidate after event selection against a background of 
$0.28^{+1.57}_{-0.14}$. A 95\%CL limit of 0.71~pb was set on the $t\bar t Z$ production (which compares with a SM prediction of 0.14~pb). 
CMS~\cite{CMS-PAS-TOP-12-014} searched, in two independent channels (three and two same charge leptons), for $t\bar t V (V=W,Z)$. While the $t\bar t V$ 
combined cross section was measured to be $\sigma(t\bar t V) = 0.51 ^{+0.15}_{-0.13}(stat) ^{+0.05}_{-0.04}(syst)$ ~pb at 7~TeV, the $t\bar t Z$ cross 
section (obtained from the trilepton channel) was $\sigma(t\bar t Z) = 0.30 ^{+0.14}_{-0.11}(stat) ^{+0.04}_{-0.02}(syst)$ ~pb 
(Figure~\ref{fig:ttGaugeBosons}).

%%%%%%%%%%%%%%%%%%%%%%%%%%%%%%%%%%%%%%%%%%%%%%%%%%%%%%%%%%%%%%%%%%%%%%%%%%%
%%%%%%%%%%%%%%%%%%%%%%%%%%%%%%%%%%%%%%%%%%%%%%%%%%%%%%%%%%%%%%%%%%%%%%%%%%%

Both ATLAS~\cite{ATLAS-CONF-2012-135} and CMS~\cite{CMS-PAS-HIG-12-025} searched for $t\bar t H$ production using data collected at 7~TeV with a total integrated 
luminosity of 5~fb$^{-1}$. While ATLAS studied the single lepton channel, where the Higgs boson is assumed to decay to a pair of $b$-quarks, CMS looked into both 
the single and dilepton channels. As no evidence for signal was found, combined upper limits, at 95\% CL, were set on cross section times branching ratio by ATLAS 
and CMS to $\sigma_{t\bar t H}\times Br(H\to b \bar b)\le$13.1~pb (10.5~pb expected), $\sigma_{t\bar t H}\times Br(H\to b \bar b)\le$3.8~pb (4.6~pb expected), 
respectively.

\section{Top quark decays beyond the SM}

In the pursuit of new physics in the top quark sector, the decays play a fundamental role. Not only they may reveal violation of symmetries naturaly observed in 
the $SU(3) \times SU(2) \times U(1)$ gauge theory of strong and electroweak interactions (which do not require introducing global separate conservation laws), as 
typically happens for baryon or lepton number conservation, but also the presence of new Physics Beyond the SM, as predicted for instance by the Minimal 
Supersymmetric Model (MSSM), could manifest itself. It could even happen that decays with very low SM branching ratios, like Flavor Changing Neutral Currents 
(FCNC), non visible at tree level and very much suppressed at one loop due to the GIM mechanism in the SM ($Br < $ 10$^{-12}$), could be experimentally detected. 
In addition to the theorectical motivation to search for these decays, the stronger motivation is perhaps more of experimental nature: to look for signs of new 
physics where the SM expects essentially no contribution.

Using 5~fb$^{-1}$ of data collected at 7~TeV, CMS~\cite{CMS-PAS-B2G-12-002} looked for the decay of top quarks that violate baryon number conservation ($t\to \bar U 
\bar D L^+$). $U, D$ and $L$ play the role of an up quark, down quark and charged lepton, respectively. The top quark decay is expected to produce one isolated electron or 
muon, two jets, and no neutrino in the final state. As no significant excess of events was found, an upper limit at 95\% CL was set on the branching ratio of the 
baryon number violating top quark decay to be 0.0067.

Both CMS~\cite{arXiv:1205.5736} and ATLAS~\cite{arXiv:1204.2760} looked for MSSM decays of top quarks to charged Higgs ($t\to H^+b$) when 
$m_{H^\pm} < m_t$. If $\tan \beta$ is particularly suited, it may happen the dominant decay mode of the charged Higgs is through the decay $H^+\to c\bar 
s$. As no sign of new physics was found, best 95\% CL upper limits were set on the branching fractions $Br(t\to H^+b)$ in the range 2-4\% for Higgs 
boson masses between 80 and 160~GeV and assuming $Br(H^+ \to \tau^+ \nu_\tau)=1$.

In the SM, FCNC decays ($t\to qX, X=\gamma, Z, g, H$) are forbidden at tree level and much smaller than the dominant decay channel. 
ATLAS~\cite{arXiv:1206.0257} and CMS~\cite{arXiv:1208.0957} searched for signs of $t\to qZ$ and, as no new physics was found, observed 
limits at 95\% CL were set on $Br(t\to qZ)\le$ 0.73\% (ATLAS) and $Br(t\to qZ)\le$ 0.27\% (CMS).

\section{Conclusions}

Top quark studies are well under way at the LHC. Although no new physics was seen yet in the top quark physics, the performance of the LHC and the ATLAS and CMS 
experiments has been remarkable. The LHC is entering a new era of precision studies for the top quark physics. 

The work of A.Onofre was supported by Funda\c{c}\~ao para a Ci\^encia e Tecnologia under project CERN/FP/123619/2011.

\end{document}